\documentclass[]{interact}
\usepackage{amssymb}
\usepackage{graphicx}
\usepackage{marginnote}
\usepackage{url}
\usepackage[caption=false]{subfig}

\graphicspath{{./figures/}}

\begin{document}

\title{Demonstration and interpretation of ``scutoid'' cells in a quasi-2D soap froth}

\author{
  \name{
    A.~Mughal$^a$,
    S.J.~Cox$^a$,
    D.~Weaire$^b$,
    S.R. Burke$^b$  and
    S.~Hutzler$^b$$^\ast$\thanks{$^\ast$Corresponding author. Email: stefan.hutzler@tcd.ie}
  }
  \affil{
    $^a$Department of Mathematics, Aberystwyth University,
Aberystwyth, Ceredigion, SY23 3BZ;\\
    $^b$School of Physics, Trinity College Dublin, The University of Dublin, Ireland
  }
}


\maketitle

\begin{abstract}
Recently a novel type of epithelial cell has been discovered and dubbed the
``scutoid''. It is induced by curvature of the bounding surfaces. We show
by simulations and experiments that such cells are to be found in a dry foam subjected to this boundary condition.
\end{abstract}


\section{Introduction}

Recently G{\'o}mez-G{\'a}lvez {\em et al.} \cite{gomez2018scutoids} have described epithelial cells of a previously unreported form which they have called \emph{scutoid}; they appear when the bounding surfaces are \emph{curved}. The distinguishing feature of such a cell is a triangular face attached to one of the bounding surfaces. Here we offer a simple illustration of this phenomenon, which is derived from the physics of foams \cite{weaire2001physics}, consisting of a computer simulation together with preliminary experimental observations.

In an ideal dry foam, bubbles enclose gas (which is treated as incompressible) and the energy is proportional to their total surface area. Alternatively, the soap films may be considered to be in equilibrium under a constant surface tension and the gas pressure of the neighbouring cells. Plateau's rules \cite{plateau1873statique}, more than a century old, place restrictions on the topology of a \emph{dry} foam (one of negligible liquid content), which is the only case considered here. 

From the earliest intrusion of physics into biology, this elementary soap froth model has attracted attention to account for the shape and development of cells \cite{thompson1942growth, dormer1980fundamental}. More sophisticated attempts to adopt it to that purpose persist today \cite{merks2005cell, bi2014energy, graner2017forms}. In the present context we show that the model largely accounts for the appearance of scutoids, in very simple and semi-quantitative terms, broadly consistent with the description in the original paper \cite{gomez2018scutoids}.

\section{Topology of dry foams}

The relevance of foams to biology is apparent from the pioneering work of
the botanist Edwin Matzke \cite{matzke1946three}. Inspired by the resemblance in shape between bubbles in foam and cells in tissues, Matzke sought to understand the forces that may be common to both. His approach was to painstakingly and exhaustively catalogue bubble shapes observed in a dry monodisperse foam, confined within a cylindrical jar. Matzke distinguished between peripheral bubbles (i.e. bubbles in contact with the walls of the cylindrical jar) and central bubbles (i.e. bubbles inside the bulk foam). Amongst the peripheral bubbles are listed 
two scutoids: the $(1,3,3,1)$ (see Figure 9-8 of \cite{matzke1946three}) 
and $(1,4,2,1,1)$ polyhedrons (as identified in Matzke's notation). Not a single triangular face was found amongst the central (i.e. bulk) bubbles.

\section{Quasi-2D foam sandwich}

\begin{figure} 
\begin{center}
\centering
\includegraphics[width=0.75\columnwidth ]{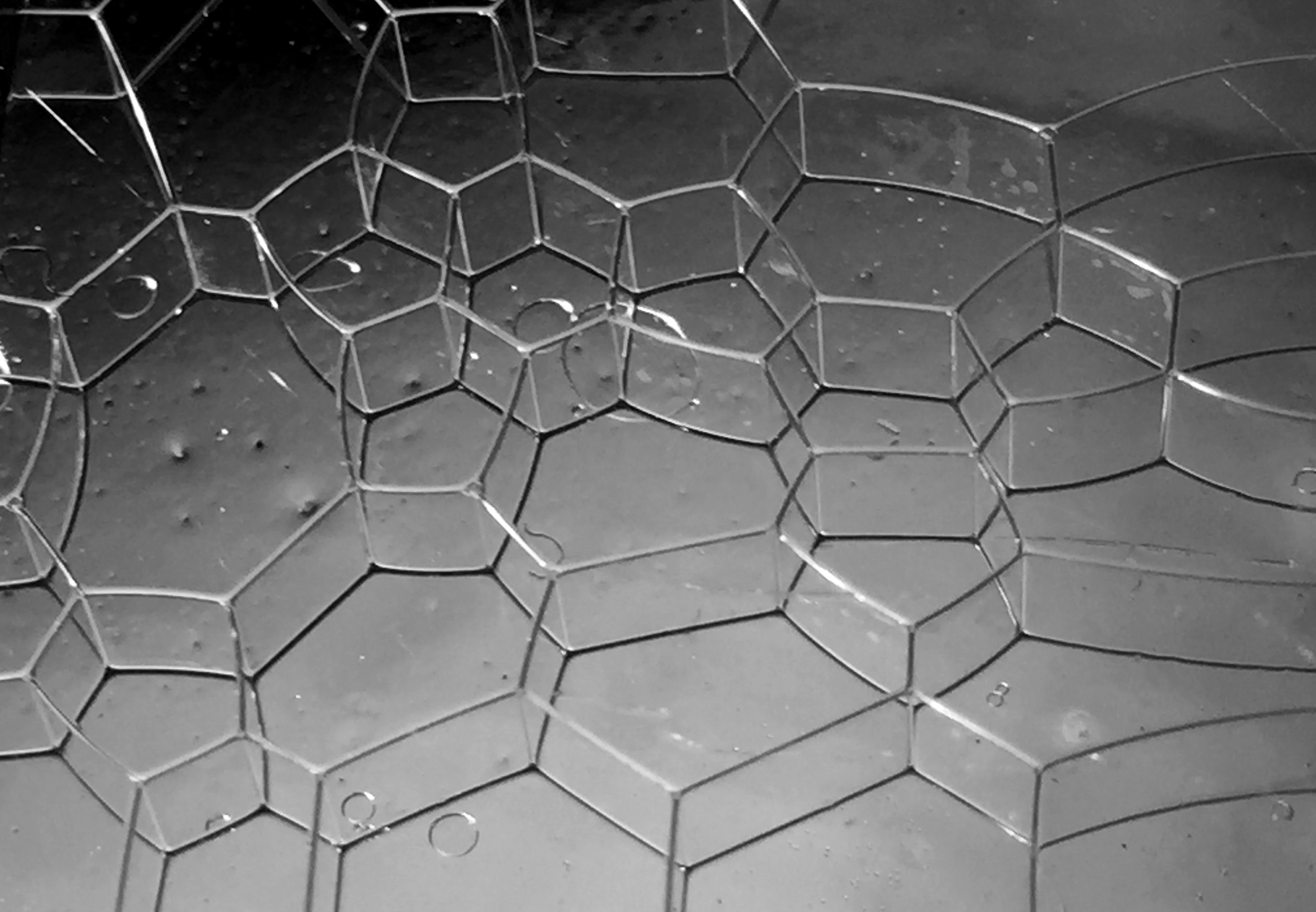}
\caption{A quasi-2D foam showing a single layer of bubbles confined 
between two flat parallel glass plates (plate
separation 8mm, average bubble diameter 2-3cm).  Internal films meet the plates at right
angles. The polygonal cells on both glass plates are identical.}
\label{quasi}
\end{center}
\end{figure}

Cyril Stanley Smith \cite{Smith52} first introduced the experimental
quasi-2D foam that is formed between two glass plates. The plates are close
enough together that all bubble cells span both boundaries, so that there
are no internal bubbles and the internal soap
films meet the glass plates at right angles (see Figure \ref{quasi}). The
quasi 2D foam between flat parallel plates is often taken as the
experimental counterpart of the ideal 2D foam - which consists of polygonal
2D cells, with (in general curved) edges meeting three at a time (only) at
$120^o$. Such a finite foam sandwich presents \emph{two} such patterns on
its two boundaries, and indeed on any plane taken parallel to them.
However, if the plate separation is increased, this structure is overtaken
by an instability, described and analysed by Cox {\em et al.}\cite{cox2002transition}, in which individual cells cease to span the two plates. This instability is not directly relevant to scutoid formation but places limitations on experiment and theory.

\begin{figure} 
\begin{center}
\centering
\includegraphics[width=0.8\columnwidth ]{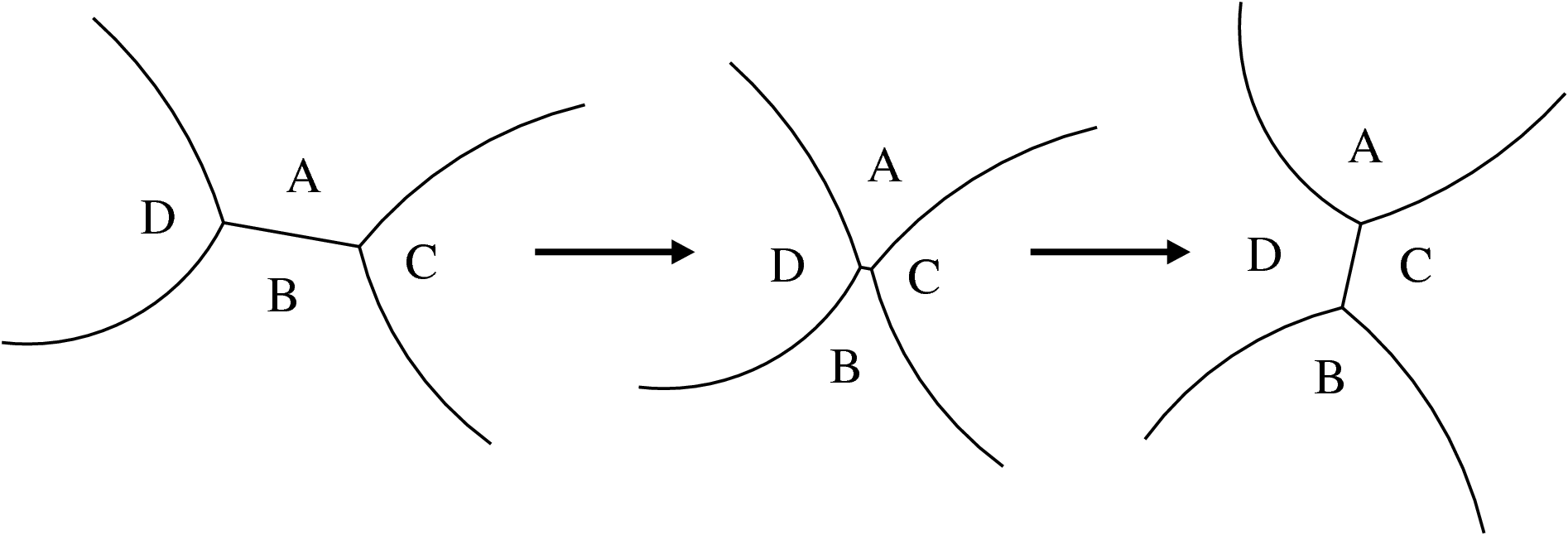}
\caption{A schematic of a T1 transition in an ideal 2D foam \cite{weaire1984soap}. The edge shared between bubbles A and B gradually shrinks and vanishes, the resulting fourfold vertex is in violation of Plateau's laws and the system transitions to a new arrangement. As a result, bubbles A and B are no longer neighbours, while C and D (which were previously unconnected) now share a boundary.}
\label{T1}
\end{center}
\end{figure}

The novel element that is brought into consideration by the work of
G{\'o}mez-G{\'a}lvez {\em et al.} is the introduction of \emph{curved}
boundaries which may be represented by two concentric cylinders or a
portion thereof. While there has been some work on the effects of curvature
of one or both plates \cite{roth2012coarsening, mughal2017curvature}, it
did not address the case considered
by G{\'o}mez-G{\'a}lvez et al which consists of two concentric boundaries.
As the separation between the two cylinders is increased, the 2D patterns
on the inner and outer surfaces become distorted, the inner one being
compressed in the circumferential direction, with respect to the outer one.
Eventually, this should lead to the vanishing of a 2D cell edge, and hence
to a topological change, as in Figure \ref{T1}. This is the so-called T1
process \cite{weaire1984soap}. It necessarily entails the creation of a
\emph{scutoid} feature within the bulk of the foam (as illustrated in
sections \ref{s:simulations} and \ref{s:expts}). However, its appearance may be only transitory, as it may provoke a similar effect of the other surface, in a double-$T1$ process that restores the original columnar structure. The geometry required by Plateau's rules makes it obvious that this must be the case if the gap between the cylinders is very small. Increasing the gap is expected to allow stable scutoids to persist, provided we do not encounter the other type of instability mentioned above.

These arguments leave room for doubt as to whether such scutoid features can really be found in the foam sandwich. Both simulations and experiments, described in the following section, have yielded positive results.   

\section{Simulations}
\label{s:simulations}

\begin{figure}[h]
\begin{center}
\centering
\includegraphics[width=0.95\columnwidth ]{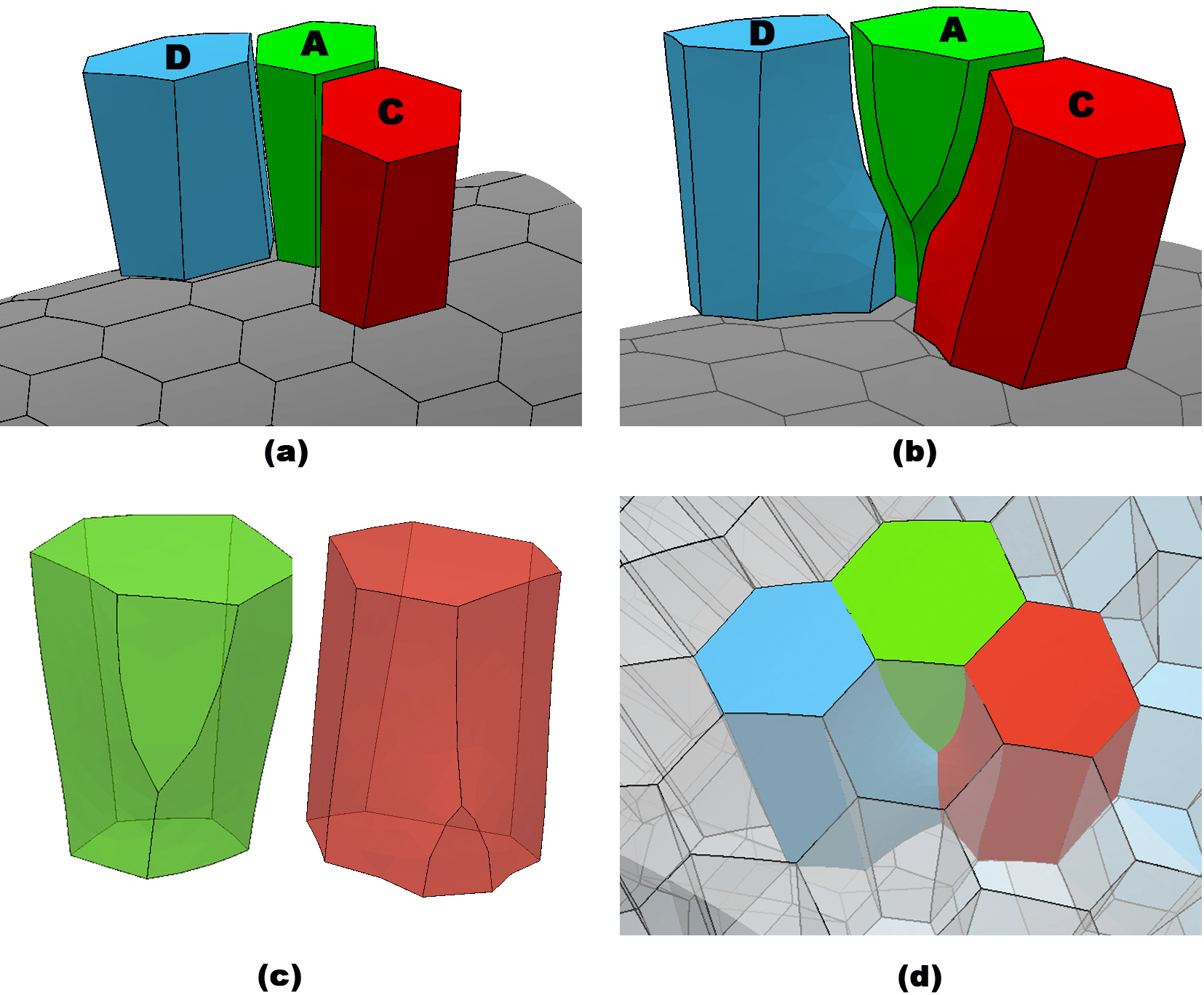}
\caption{Cells in a Surface Evolver simulation of a polydisperse foam confined between two concentric cylinders. (a) In the initial state the 2D pattern on both boundaries is purely hexagonal (only the pattern on the substrate is shown). Red and blue bubbles are not in contact while the green bubble is in contact with a fourth neighbouring bubble (not shown for clarity). (b) The foam after a T1 transition on the substrate, resulting in four stable scutoid cells. The pattern on the substrate contains two five-sided and two seven-sided regions, while the pattern on the superstrate remains purely hexagonal. The cells are shown slightly separated for clarity. (c) The two types of scutoids cells (pentagonal and heptagonal) are shown separately. (d) A combined view showing the scutoids and the surrounding foam cells.
}
\label{SEscutoid}
\end{center}
\end{figure}

As in the simulations of \cite{gomez2018scutoids}, we start from a Voronoi partition of the gap between two concentric cylinders, to give a collection of hexagonal prismatic cells. This structure is imported into the Surface Evolver software \cite{brakke1992surface}, which permits the minimization of surface energy (here equivalent to surface area, as in the ideal foam model) subject to fixed cell volumes. We employ a periodic boundary condition in the direction of the axis of the cylinders to reduce the effect of the finite size of the simulation. Cell volumes are assigned fixed values within a restricted range so that the initial structure is polydisperse but still hexagonal. In the example shown in Fig \ref{SEscutoid}a, the cylinder has axis length 5.2 units, the cylinder radii are 2.8 and 4.3 units and there are 144 cells. To allow the cell walls to develop realistic curvature, we tessellate each face with small triangles and perform a standard Surface Evolver minimization of the surface area. 

In this preliminary exploration topological changes were triggered using the Surface Evolver software. A number of stable scutoids were identified of which one example is shown in Figure \ref{SEscutoid}. In the continuation of this work we expect to map out the parameter space in which such stable scutoids are to be found.

\section{Experiments with soap bubbles}
\label{s:expts}

\begin{figure} 
\begin{center}
\centering
\includegraphics[width=0.8\columnwidth ]{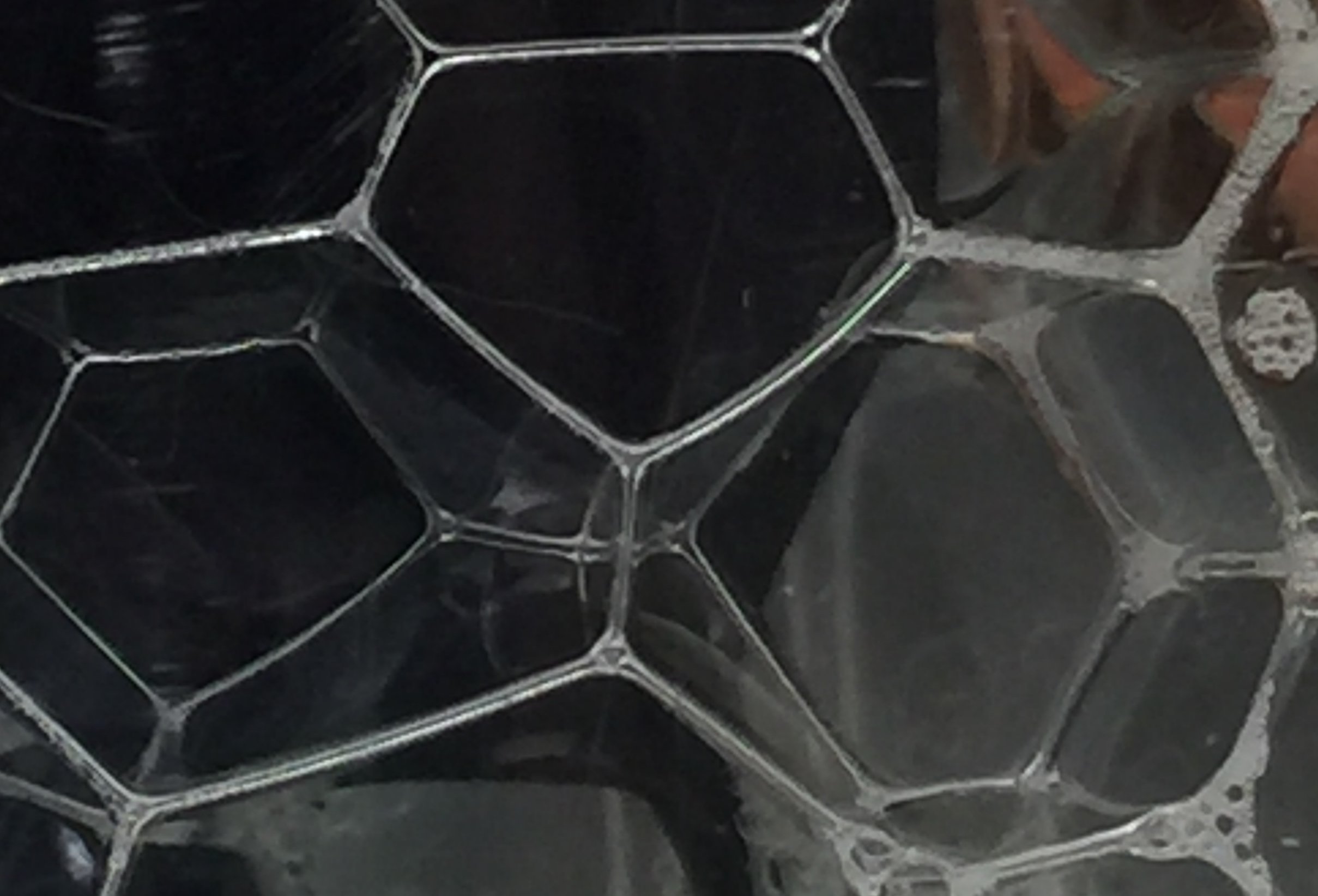}
\caption{Photograph of scutoids in a quasi-2d foam sandwich. The bubble on the left
features a hexagon in contact with the outer cylinder and a pentagon in
contact with the inner cylinder while the bubble on the right shows a
heptagon on the outer and a hexagon on the inner cylinder. Also visible is
the small triangular face separating these two bubbles. (diameter of inner 
cylinder
$21$mm, internal diameter of hollow outer cylinder $39$mm, spacing about $7$mm, approximate equivalent sphere diameter of the bubbles 8 mm.)
}
\label{scutoid-photo}
\end{center}
\end{figure}

We performed preliminary experiments with soap bubbles between curved
surfaces, using a  glass cylinder of
diameter $21$mm as a substrate and a hollow half cylinder (made from perspex) 
with inner diameter $39$mm as a superstrate. 
The bubbles (approximate equivalent
sphere diameter 8 mm) were produced using a simple aquarium pump with
flow control and commercial dish-washing solution. Rather than placing the
two cylinders upright into the vessel containing the solution we placed them on
their long axis, creating an approximately 7mm wide gap between them, which
was initially about
half-way filled with liquid. We then used a syringe needle attached to the
pump to blow gas into this gap, leading to the formation of a quasi-2D foam
sandwich. By reducing the water level we arrive at bubbles which are in
contact with both cylinder surfaces, some of them forming scutoids, see
Figure \ref{scutoid-photo}. The
present process involves a measure of trial and error: repeated raising and
lowering the water level allows for repeated bubble rearrangements which
increases the chance of finding  scutoids.

\section{Conclusion}

Both simulation and experiment have confirmed that stable scutoid configurations are to be found in a dry foam sandwich between cylindrically curved faces. It remains for future work to identify the conditions for this in terms of geometrical parameters. 

The foam model is well established in the description of biological cells and the processes by which they change their arrangements, but is at best a rough first approximation. In the present case we have noted that epithelial cells may be relatively elongated. If greater realism is called for, further energy terms may be added, stiffening the cell walls. 

\section{Acknowledgements}

This research was supported in part by a research grant
from Science Foundation Ireland (SFI) under grant number 13/IA/1926.
A. Mughal acknowledges the Trinity College Dublin Visiting Professorships and 
Fellowships Benefaction Fund.
We thank B. Haffner for providing the photograph of Figure \ref{quasi}.

\bibliographystyle{tfq}

\begin{thebibliography}{10}
\newcommand{\printfirst}[2]{#1}
\newcommand{\switchargs}[2]{#2#1}
\providecommand{\url}[1]{\normalfont{#1}}
\providecommand{\urlprefix}{Available at }

\bibitem{gomez2018scutoids}
P. G{\'o}mez-G{\'a}lvez, P. Vicente-Munuera, A. Tagua, C. Forja, A.M. Castro,
  M. Letr{\'a}n, A. Valencia-Exp{\'o}sito, C. Grima, M. Berm{\'u}dez-Gallardo,
  {\'O}. Serrano-P{\'e}rez-Higueras, F. Cavodeassi, S. Sotillos, M.D.
  Mart{\'i}n-Bermudo, A. M{\'a}rquez, J. Buceta, and L.M. Escudero,
  Nat. Commun. 9 (2018) p. 2960.

\bibitem{weaire2001physics}
D. Weaire and S. Hutzler, \emph{The physics of foams}, Clarendon Press, Oxford,
  1999.

\bibitem{plateau1873statique}
J.A.F. Plateau, \emph{Statique Exp\'erimentale et Th\'eorique des Liquides
  soumis aux seules Forces Mol\'eculaires}, Gauthier-Villars, Paris, 1873.

\bibitem{thompson1942growth}
D.W. Thompson, \emph{On growth and form}, Cambridge University Press, 1917.

\bibitem{dormer1980fundamental}
K. Dormer, \emph{Fundamental tissue geometry for biologists}, Cambridge
  University Press, Cambridge, 1980.

\bibitem{merks2005cell}
R.M. Merks and J.A. Glazier, 
Phys. A  352 (2005) p. 113.

\bibitem{bi2014energy}
D. Bi, J.H. Lopez, J. Schwarz, and M.L. Manning, 
Soft Matter 10 (2014) p. 1885.

\bibitem{graner2017forms}
F. Graner and D. Riveline, 
Development  144 (2017) p. 4226.

\bibitem{matzke1946three}
E.B. Matzke, 
  Am. J. Bot. 33 (1946) p. 58.

\bibitem{Smith52}
C. Smith, 
Metal Interfaces (ASM Cleveland)  (1952) p. 65.

\bibitem{cox2002transition}
S. Cox, D. Weaire, and M.F. Vaz, 
Eur. Phys. J. E 7 (2002) p. 311.

\bibitem{roth2012coarsening}
A. Roth, C. Jones, and D.J. Durian, 
Phy. Rev.  E 86 (2012) p. 021402.

\bibitem{mughal2017curvature}
A. Mughal, S. Cox, and G. Schr{\"o}der-Turk, 
Interface focus 7 (2017) p. 20160106.

\bibitem{weaire1984soap}
D. Weaire and N. Rivier, 
Contemp. Phys. 25 (1984) p. 59.

\bibitem{brakke1992surface}
K.A. Brakke, 
Exp. Math. 1 (1992) p.  141.

\end{thebibliography}

\end{document}